# Modeling and scaling spontaneous imbibition with generalized fractional flow theory and non-Boltzmann transformation

Shaluka Senevirathna[1], Anna Zemlyanova[2], Shaina A. Kelly[3], Qinhong Hu[4], Yong Zhang[5] and Behzad Ghanbarian[6,7,8*]

[1] Porous Media Research Lab, Department of Geology, Kansas State University, Manhattan KS 66506 USA

[2] Department of Mathematics, Kansas State University, Manhattan KS 66506 USA

[3] Department of Earth and Environmental Engineering, Columbia University, New York, NY 10027 USA

[4] State Key Laboratory of Deep Oil and Gas, China University of Petroleum (East China), P.R. China

[5] Department of Geological Sciences, University of Alabama, Tuscaloosa, AL 35487 USA

[6] Department of Earth and Environmental Sciences, University of Texas at Arlington, Arlington TX 76019, United States

[7] Department of Civil Engineering, University of Texas at Arlington, Arlington TX 76019, United States

[8] Division of Data Science, College of Science, University of Texas at Arlington, Arlington TX 76019, United States

[*] Corresponding author's Email address: ghanbarianb@uta.edu

**Abstract**

Spontaneous imbibition (SI) is a process by which liquid is drawn into partially saturated porous media by capillary forces, relevant for subsurface processes like underground fluid storage and withdrawal. Accurate modeling and scaling of counter-current SI have long been challenging. In this study, we proposed a generalized fractional flow theory (GFFT) using the Hausdorff fractal derivative, combined with non-Boltzmann scaling. The model links imbibition distance to time through the power law exponent $\alpha/2$, where $\alpha$ is the fractal index ($0 < \alpha < 2$ in this study). We applied the GFFT to various experimental and stimulated datasets of both porous and fractured media, finding that $\alpha$ varied with factors such as contact angle (of the imbibing fluid), dynamic viscosity, pore structure, and fracture properties. By analyzing SI data from sandstones, diatomite, carbonate, and synthetic porous media, we demonstrated that the non-Boltzmann scaling provided a better collapse of the SI data than the traditional Boltzmann approach ($\alpha = 1$), with $\alpha$ values ranging from 0.88 to 1.54. These deviations illustrate the model's adaptability to different porous materials. Using the GFFT, we expect to better predict fluid imbibition rates when properties like porosity, permeability, initial and maximum saturations, viscosity, and wettability are known, offering a more accurate alternative to traditional models.

1.  **Introduction**

Spontaneous imbibition (SI), a crucial process through which a wetting phase (e.g., water) imbibes spontaneously into a porous medium and displaces a non-wetting phase (e.g., air), is driven by capillary forces and tempered by viscous forces [1]. A real-world application of SI is $CO_2$ sequestration on large scales, a practical strategy to reduce atmospheric greenhouse gases and combat climate change. During underground fluid sequestration, $CO_2$, injected under supercritical conditions as a possible wetting phase, enters subsurface formations. SI rate and extent are primarily influenced by permeability, initial water content, viscosity, pore structure, and rock wettability [2]. In gas-liquid systems, gas is commonly the non-wetting phase, while in liquid-liquid systems (e.g., water-oil), the non-wetting phase is a function of material hydrophilicity.

There are two main types of SI processes depending on system boundary conditions: (1) counter-current and (2) co-current [3]. In counter-current SI, non-wetting and wetting fluids flow in opposite directions, while in co-current SI, both fluids move in the same direction. Co-current SI scenarios occur in capillary tube-style experiments [4] and in porous materials where only one end of the domain is exposed to wetting fluid. Counter-current SI scenarios occur when a sample is exposed to the wetting fluid on all sides, and/or when there is a boundary condition that creates a backpressure of the non-wetting fluid. Counter-current SI is considered the predominant driving mechanism among the two modes in geologic porous media, as evidenced by extensive experimental and computational studies on this topic [3,5,6]. In addition, counter-current SI is more likely to occur in tight (low permeability) rock matrices or dual-porosity systems, where counter-current SI occurs at the matrix-fracture interface. This was evidenced at Ekofisk tight chalk field formations where oil was displaced into the fracture system by water imbibed into the matrix [7].

## 1.1. Modeling spontaneous imbibition

Since mid-1950s, the literature on SI is vast and extensive [8]. Experimental approaches [9,10], theoretical models [11,12], numerical simulations [6,13], and artificial intelligence-based models [14,15] have been widely applied to study SI in porous media, particularly in tight rocks [16,17] and fracture networks [18]. For a review of SI equations and methods, see Abd et al. [5].

Among theoretical approaches, several models were developed based on "bundle of capillary tubes" approach, an oversimplifying idealization of a porous medium, where the network of pores is replaced by non-interconnected tortuous tubes spanning the entire medium [22]. For instance, Cai et al. [11] proposed a model for the SI using the bundle of capillary method. Similarly, Shi et al. [19] developed a theoretical model that excludes hydrostatic pressure and a semi-analytical model that includes it, both using the bundle of capillary tubes combined with fractal theory. However, those authors did not validate their derivations through either laboratory or numerical experiments.

## 1.2. Scaling analysis of SI

Scaling analysis of counter-current SI data for various porous media has been a long-standing challenge. It requires a thorough understanding of SI mechanisms, the ability to capture fluid-rock characteristics affecting the imbibition process across different rock types, and the incorporation of these factors into a mathematical framework.

In the literature, various approaches have been proposed for the SI scaling analysis through which normalized imbibed volume is linked to some dimensionless time [3]. In one of the first attempts, Mattax and Kyte [20] introduced a mathematical framework that integrated both gravitational and capillary forces based on the model of Rapoport [21] and proposed a dimensionless time. Their model considered capillarity, gravitational effects on fluid saturation, viscosity, and core dimensions. Later, Ma et al. [23] proposed a different scaling law by introducing a new characteristic length to accommodate various boundary conditions. Their model, however, works well only for rocks with similar wettability, relative permeabilities, and oil/water/rock systems [24]. Zhang et al. [25] showed that the Ma et al.'s [23] model failed to scale SI data in gas/liquid/rock systems. To address this, Li and Horne [26] developed a new model based on solution to recovery by spontaneous water imbibition, which they validated in gas/water/rock systems using different rock types (Berea sandstone, chalk, and graywacke) under varying initial water saturations. Their model was later generalized to scale both co-current and counter-current SI data in gas/liquid/rock and oil/water/rock systems [24].

While the literature presents a variety of SI models, many have inherent limitations. For example, several models fail to account for certain critical factors, such as capillary pressure curves, relative permeability, and wettability, limiting their broader applicability. Capturing salient properties that control SI is important for accurate modeling; failure to do so reduces model accuracy. Recognizing these limitations, researchers have endeavored to integrate capillarity and permeability effects or introduce supplementary assumptions to craft more comprehensive models [1]. Despite these efforts, the universal applicability of these models across diverse porous media remains a challenge.

One of the theoretical frameworks proposed to scale counter-current SI is based on fractional flow theory, first developed by Buckley and Leverett [27]. Schmid and Geiger [1] applied fractional flow theory in combination with the Boltzmann transformation ($x =$ $\lambda t^{0.5}$ in which $x$ is the distance, $t$ is the time, and $\lambda$ is a numerical prefactor) to model SI and its scaling. To test their approach, Schmid and Geiger [1] evaluated 42 datasets from water-oil and water-air experiments, where water served as the wetting phase.

**1.3. Deviation from Boltzmann scaling**

Although Boltzmann scaling has been widely applied in the literature to characterize SI, several studies have shown deviations from the exponent of 0.5 in the Boltzmann transformation due to factors, such as complex, tight pore structures and nano-confinement effects. For instance, Hu et al. [28] reported three types of imbibition slopes that were different from 0.5, further observed by Gao and Hu [29] in other types of rocks and conditions. Hu et al. [28] performed experiments and network simulations to show that porous media with low pore connectivity exhibit imbibition slopes as low as 0.26.

In another study, Kelly [30] reported experimental results based on nanofluidic chips with parallel nanochannels of varied cross-sectional sizes and used image and data analysis schemes to show the deviation of imbibition time exponents from 0.5. Kelly et al. [32] also found that imbibition in silica nanochannels was at least five times slower than that predicted by the Washburn equation for polar fluids, such as water and isopropanol. These controlled experiments indicated that special considerations may be needed for imbibition dynamics in nano-porous geologic media, such as shales and igneous rocks (e.g., basalts and dunites).

By means of numerical simulations, Ning et al. [31] studied imbibition in fractured Barnett Shale and showed that the imbibition rate was influenced by fracture parameters and connectivity among grid cells. They also found deviations from Boltzmann scaling and reported exponents different from 0.5. Their results suggested that the imbibition process is affected by various factors, including fluid viscosity and temperature, material wettability, fracture distribution, and pore structure as well as boundary conditions such as initial saturation and maximum saturation.

## 2. Objectives

There are both experimental measurements and numerical simulations in the literature that strongly suggest deviations from Boltzmann scaling at both pore [32-34] and core [28,35-37] scales. In addition, although various approaches have been proposed to scale counter-current SI data, normalized imbibed volume versus some dimensionless time still does not satisfactorily collapse onto a universal curve [1]. The main objectives of this study, therefore, are to: (1) develop a new model for counter-current SI by generalizing fractional flow theory using fractal derivatives and non-Boltzmann transformation; (2) present a novel scaling law to better collapse SI data than previous models; and (3) evaluate the proposed approaches using experimental and numerical data for different types of porous media.

In the following sections, we focus on counter-current SI, which, for simplicity, we refer to as SI.

## 3. Theory

### 3.1. Fractional flow theory

Buckley and Leverett [27] presented the fractional flow theory to describe the physics of immiscible displacements in porous media. After their pioneering work,

McWhorter and Sunada [38] provided exact quasi-analytical solutions for both one-dimensional and radial flow of two incompressible fluids and calculated their fractional-flow functions for the first time. The partial differential equation for the one-dimensional horizontal flow of two immiscible and incompressible fluids is [38]

$$\phi \frac{\partial S_w}{\partial t} = -q_t \frac{\Delta f(S_w)}{\Delta S_w} \frac{\partial S_w}{\partial x} + \frac{\partial}{\partial x}\left[D(S_w) \frac{\partial S_w}{\partial x}\right] \qquad (1)$$

where $S_w$ is the wetting-phase saturation, $t$ is the time, $q_t$ is the total volume flux, $\phi$ is the porosity, $D$ is the diffusivity coefficient, and $f$ is the fractional flow function which describes the flow rate of a fluid relative to the total flow rate. $D$ and $f$ are expressed as follows [38]

$$f(S_w) = \left(1 + \frac{k_{nw}\mu_w}{k_w\mu_{nw}}\right)^{-1} \qquad (2a)$$

$$D(S_w) = -\frac{k_{nw}f(S_w)}{\mu_{nw}} \frac{\Delta P_c}{\Delta S_w} \qquad (2b)$$

where $k_w$ and $k_{nw}$ are the (medium) permeabilities to the wetting and non-wetting fluids, respectively, $\mu_w$ and $\mu_{nw}$ are the wetting- and non-wetting-phase dynamic viscosities, respectively, and $P_c$ is the capillary pressure.

In the case of counter-current SI, the total flux ($q_t$) is zero because the volume flux of the wetting phase is equal and opposite to that of the non-wetting phase ($q_w = -q_{nw}$), and thus, Eq. (1) reduces to [39]

$$\phi \frac{\partial S_w}{\partial t} = \frac{\partial}{\partial x}\left[D(S_w) \frac{\partial S_w}{\partial x}\right]. \qquad (3)$$

**3.2. Generalized fractional flow theory**

Fractal derivatives have been successfully used to model transport phenomena in porous media [43] and to study two-phase flow in porous rocks [41]. Using the concept of the Hausdorff fractal derivative proposed by Chen [40], we generalized fractional flow theory by introducing the time fractal form of Eq. (3) as follows

$$\phi\varepsilon\frac{\partial S_w}{\partial t^\alpha} = \frac{\partial}{\partial x}\left[D_\alpha^F(S_w)\frac{\partial S_w}{\partial x}\right] \qquad (4)$$

where $\varepsilon$ $[T^{\alpha-1}]$ (=1) is a unit conversion factor, and $\alpha$ [dimensionless] is the order of the Hausdorff fractal derivative, as defined below [40]:

$$\frac{\partial g(t)}{\partial t^\alpha} = \lim_{t_1 \to t}\frac{g(t_1)-g(t)}{t_1^\alpha - t^\alpha},$$

where $g(t)$ is a function of the variable $t^\alpha$. This formula represents a fractal ruler in time, re-scaling the clock time (either shortening or expanding) that each fluid packet spends imbibing in a complex medium. The index $\alpha$, which is an upscaling parameter, captures the combined effects of the medium, fluids, and their interactions on SI dynamics.

Hereafter, we refer to Eq. (4) as the generalized fractional flow theory (GFFT). Theoretically, the exponent $\alpha$ ranges between 0 and 2 [43]. When $0 < \alpha < 1$, the GFFT model captures non-Boltzmann scaling where the wetting front moves slower than the square root of time, indicating sub-diffusive imbibition. When $1 < \alpha < 2$, it describes faster movement, indicating super-diffusive imbibition. When $\alpha = 1$, the GFFT, Eq. (4), simplifies to Eq. (3) for Boltzmann scaling.

Eq. (4) is similar in form to the time-fractal Richards' equation first proposed by Sun et al. [43]:

$$\frac{\partial S_w}{\partial t^\alpha} = \frac{\partial}{\partial x}\left[D_\alpha^R(S_w)\frac{\partial S_w}{\partial x}\right] \qquad (5)$$

in which $D_\alpha^R(S_w)$ is defined as

$$D_\alpha^R(S_w) = k_w \frac{\Delta P_c}{\Delta S_w} \qquad (6)$$

One of the differences between Eqs. (4) and (5) is the inclusion of porosity ($\phi$) in the GFFT Eq. (4), which is absent in Eq. (5). Note that porosity $\phi$ is treated as a constant here, as it represents the material's available pore space for fluid flow, regardless of which phase occupies that space. Furthermore, $D_\alpha^F(S_w)$ in Eq. (4) follows the definition given in Eq. (2b), which is different from $D_\alpha^R(S_w)$ in the time-fractal Richards' equation defined in Eq. (6). In fact, $D_\alpha^F(S_w)$ in Eq. (2b) is a non-monotonic function of $S_w$ (see Fig. 4b in McWhorter and Sunada [38]), whereas $D_\alpha^R(S_w)$ in Eq. (5) has been assumed to monotonically increase with increasing $S_w$ [36].

To model counter-current SI, one should combine the GFFT with the non-Boltzmann transformation, $x = \lambda t^{\frac{\alpha}{2}}$, where $x$ is the distance that the saturated front travels in the medium, $t$ is time, $0 < \alpha < 2$ is the non-Boltzmann constant, and $\lambda$ is a numerical pre-factor. Using the relationship $\lambda = xt^{-\alpha/2}$ and the chain rule gives the following equations:

$$\frac{\partial S_w}{\partial t^\alpha} = \frac{dS_w}{d\lambda}\frac{\partial \lambda}{\partial t^\alpha} = -\frac{dS_w}{d\lambda}\frac{\lambda}{2t^\alpha} \tag{7}$$

$$\frac{\partial S_w}{\partial x} = \frac{dS_w}{d\lambda}\frac{\partial \lambda}{\partial x} = \frac{dS_w}{d\lambda}\frac{1}{t^{\alpha/2}}. \tag{8}$$

Substituting Eq. (7) and Eq. (8) into Eq. (4) produces a non-linear, second-order ordinary differential equation for the function $S_w = S_w(\lambda)$:

$$-\phi\varepsilon\frac{\lambda}{2}\frac{dS_w}{d\lambda} = \frac{d}{d\lambda}\left[D_\alpha^F(S_w)\frac{dS_w}{d\lambda}\right]. \tag{9}$$

Introducing new unknown functions as $z(\lambda) = S_w(\lambda)$ and $w(\lambda) = S_w'(\lambda)$, Eq. (9) can be reduced to a pair of first-order ordinary differential equations as follows:

$$z' = w \tag{10}$$

$$w' = -\frac{D_\alpha^{F\prime}(z)w^2 + \phi w\lambda/2}{D_\alpha^F(z)} \tag{11}$$

In a medium with a constant initial water saturation, one may set the initial condition $S_w(t = 0) = S_i$. Eqs. (10) and (11) are subject to the following boundary conditions for the function $z(\lambda)$:

$$S_w(x, t = 0) = S_w(\infty, t) = S_i \qquad z(\infty) = S_i \tag{12a}$$

$$S_w(x = 0, t) = S_0 \qquad z(0) = S_0 \tag{12b}$$

We should note that the proposed solution, Eqs. (10)-(11), is different from those developed by Pachepsky et al. [42], Sun et al. [43], and Li et al. [41].

### 3.3. Scaling counter-current spontaneous imbibition

Based on the non-Boltzmann transformation, we propose a dimensionless time, $t_d$, as follows (see Appendix A)

$$t_d = \left(\frac{2A}{\phi\sqrt{\varepsilon}L_c}\right)^{2/\alpha} t \tag{13}$$

We should point out that the proposed scaling approach reduces to that of Schmid and Geiger [1] when $\alpha = 1$. Recall that $\varepsilon = 1 \; [T^{\alpha-1}]$ is a unit conversion factor.

The parameter $A \; [LT^{-0.5}]$ in Eq. (13) can be determined from [38]

$$A^2 = \frac{\phi}{2} \int_{S_i}^{S_0} \frac{(S_w - S_i) D(S_w)}{F(S_w)} dS_w \tag{14a}$$

in which,

$$F(S_w) = 1 - \frac{\int_{S_w}^{S_0} \frac{(\beta - S_w) D(S_w)}{F(S_w)} d\beta}{\int_{S_i}^{S_0} \frac{(S_w - S_i) D(S_w)}{F(S_w)} dS_w} \tag{14b}$$

In Eq. (13), the characteristic length $L_c \; [L]$, incorporating the boundaries of fluid imbibition in the matrix under consideration, can be defined as follows [23]:

$$L_c = \sqrt{\frac{V_b}{\sum_{i=1}^{n} A_i / l_{a_i}}} \tag{15}$$

where $V_b$ $[L^3]$ is the porous medium bulk volume, $A_i$ $[L^2]$ is the area exposed to the imbibition in the $i^{th}$ direction, and $l_{ai}$ $[L]$ is the distance the imbibition front travels towards the no-flow boundary.

## 4. Materials and Methods

### 4.1. Modeling counter-current SI

We compared the proposed GFFT model, Eq. (4), with experiments and simulations in different types of porous media to evaluate its accuracy and applicability. In this work, with the utilization of the new model for SI based on the GFFT (Section 3), a more precise prediction of liquid imbibition into a porous medium is expected when key characteristics such as porosity, permeability, initial and maximum saturations, viscosity, and wettability are known. In this section, we briefly describe a non-exhaustive set of SI experiments and simulations collected from the literature. Next, we explain how the GFFT was fit to those datasets using an in-house Python code and how the exponent $\alpha$ was optimized.

#### 4.1.1. Ferguson and Gardner (1963) dataset

We analyzed the experimental data from Ferguson and Gardner [45], who measured the time needed for an infiltration water front to reach various locations along a soil column. Their experiments were conducted on a Salkum silty clay sample, with initial conditions preserved. They measured and reported soil water content as a function of time at different locations along the soil column, ranging from 3.5 cm to 27.5 cm, using gamma ray absorption equipment. In their study, soil water diffusivity was a monotonically increasing function of water content (see their Fig. 5). We, therefore, used the measured $D_\alpha^R(S_w)$, fit the solution of the fractal Richards' equation, Eq (5), to the measured water saturation curves, and optimized the non-Boltzmann transformation exponent $\alpha$ using the nonlinear least square method.

### 4.1.2. El Abd and Milczarek (2004) dataset

This dataset includes two types of building materials: (1) fired brick clay and (2) siliceous brick from El Abd and Milczarek [46]. Those authors conducted experiments on the kinetics of the wetting process in these two building materials using dynamic neutron radiography. The samples were prepared as bars, air dried, and experimentally measured for imbibition [46]. They recorded real time images, or radiograms, to determine the advancing water content variation into the media at different positions (or distances) as a function of time. We fit the solution of the time-fractal Richards' equation, Eq. (5), to the data and optimized the exponent $\alpha$. For both experiments, the power-law water diffusivity in the form of $D_\alpha^R(\theta) = C_0 \theta^n (\frac{mm^2}{s^\alpha})$, in which $\theta$ is the water content, was used as input to the fractal Richards' equation. The parameters were $C_0 = 0.075$ and $n = 1.75$ for the fired-brick clay, and $C_0 = 0.98$ and $n = 8.2$ for the siliceous brick, as reported by Sun et al. [43].

### 4.1.3. Qin et al. (2022) dataset

This dataset includes four sets of simulations on the Estaillades carbonate reported by Qin et al. [47], who investigated the effect of pore-scale heterogeneity through direct flow stimulations. The saturation-time profiles were computed for different contact angles, i.e., 20°, 40°, 60°, and 80°. To determine $D_\alpha^F(S_w)$ via Eq. (2b), we first collected three capillary pressure curves for the same rock type, reported by Bultreys et al. [48], Bultreys et al. [49], and Qin et al. [47]. We then fit the van Genuchten [50] model to the collected capillary pressure data and optimized its parameters to estimate wetting and non-wetting-phase relative permeability curves, required in Eq. (2b), using the van Genuchten-Mualem model (see Section 4.1.5 for details). Afterward, we applied the GFFT to fit the saturation-time profiles and optimized the exponent $\alpha$ for each contact angle simulation.

We also demonstrated that their saturation-time stimulations could be scaled to fall onto a single curve using the GFFT model developed in this study. Specifically, we showed that the $\alpha$ values used to scale (or collapse) the curves were close to those optimized by fitting the GFFT to the data.

### 4.1.4. Meng et al. (2022) dataset

Meng et al. [51] reported three experiments for counter-current SI in quartz sand packs. They used quartz sands with irregular shapes and relatively narrow particle size distributions. Synthetic brine ($\mu_w$ = 1 mPa s) was used as the wetting phase, while kerosene ($\mu_{nw}$ = 2.8 mPa s), white oil-15 ($\mu_{nw}$ = 25.6 mPa s), and white oil-32 ($\mu_{nw}$ = 103.4 mPa s) were used as the non-wetting phases. To calculate the $D_\alpha^F(S_w)$, we used the capillary pressure data for quartz sand reported by Ghanbarian et al. [52] and optimized the van Genuchten [50] model parameters by fitting it directly to the reported capillary pressure curve. The wetting and non-wetting relative permeabilities were then estimated using the van Genuchten-Mualem model and used to estimate $D_\alpha^F(S_w)$ via Eq. (2b). Next, we fit the GFFT, Eq. (4), to the normalized imbibed volume-time measurements and optimized the $\alpha$ value for each experiment.

In addition to modeling the Meng et al. [51] data using the GFFT, we scaled the oil recovery SI and time, showing that the normalized imbibed volume and dimensionless time data collapsed onto a single curve. Specifically, we compared the $\alpha$ values optimized by directly fitting the GFFT solution to the data with those used to collapse them. Using the Meng et al. [51] dataset, we also investigated the viscosity dependence of $\alpha$.

### 4.1.5. Fitting GFFT to data

One of the main inputs for the solution of GFFT, Eq. (4), is $D_\alpha^F(S_w)$ defined in Eq. (2b). The $D_\alpha^F(S_w)$ is calculated based on capillary pressure and relative permeability data, which are salient properties in two-phase flow studies and in describing SI. In this study, when $k_w$ and $k_{nw}$ were not available, we estimated them from the capillary pressure curve using the van Genuchten-Mualem model and the following equations [53]:

$$k_w = \left(\frac{S_w - S_{wr}}{1 - S_{wr}}\right)\left[1 - \left(1 - \frac{S_w - S_{wr}}{1 - S_{wr}}\right)^{1/m}\right]^{2m} \tag{16}$$

$$k_{nw} = \left[1 - \left(\frac{S_w - S_{wr}}{1 - S_{wr} - S_{nr}}\right)\right]\left[1 - \left(\frac{S_w - S_{wr}}{1 - S_{wr} - S_{nr}}\right)^{1/m}\right]^{2m} \tag{17}$$

where $k_w$ and $k_{nw}$ are the relative permeabilities of the wetting and non-wetting phases, respectively, $S_w$ is the wetting saturation, $S_{wr}$ is the residual wetting saturation, and $S_{nr}$ is the residual non-wetting phase saturation. The parameters $m$ and $S_{wr}$ were optimized by fitting the van Genuchten model to the capillary pressure data.

After determining the $D_\alpha^F(S_w)$, we solved the ordinary differential equations, Eqs. (10) and (11), as a boundary value problem using the given boundary conditions, by developing an in-house Python code and using the solve.bvp package. This solution allowed us to calculate the phase imbibition into the medium as a function of time.

**4.2. Scaling analysis of counter-current SI**

To evaluate the GFFT and non-Boltzmann transformation approach in the counter-current SI scaling, we collected data, including experimental measurements and numerical stimulations covering a wide range of porous materials from the literature (see Table 1). We analyzed the experimental SI data reported by Fischer et al. [54], Zhang et al. [25], Zhou et al. [55], Hamon and Vidal [56], Bourblaux and Kalaydjlan [57], and Babadagli and Hatiboglu [58], which includes 25 sandstones, 2 diatomites, and 6 synthetic porous media. We also analyzed numerical stimulation SI data reported by Qin et al. [47] and Meng et al. [51], which includes carbonates and quartz sand. To collapse the SI data into a single curve, we first normalized the imbibed volume, reported as either recovery factor or water saturation, by its maximum value (see Zhang et al. [25]) and then plotted that against the dimensionless time determined via Eq. (15) proposed in this study. We next optimized the non-Boltzmann transformation exponent $\alpha$ to obtain the best collapse in the SI data. Using the same SI data, we evaluated the scaling approach proposed by Schmid and Geiger [1] and demonstrated that incorporating the non-Boltzmann exponent resulted in a better collapse of the data. The following subsections, we briefly describe the digitized data within each dataset used for the scaling analysis of the counter-current SI in this study.

**4.2.1. Zhang et al. (1996) dataset**

Zhang et al. [25] reported experimental SI data for seven Berea sandstones. A synthetic reservoir brine and white oil were used as the wetting and non-wetting phases for the imbibition experiments, respectively. The cores were prepared with different boundary conditions: one end open (OEO), two ends open (TEO), and two ends closed (TEC) during imbibition. They were oven-dried to ensure non-saturated initial conditions, and counter-current imbibition experiments were carried out. We digitized the oil recovery and time data for the counter-current SI scaling analysis. Core dimensions and other properties summarized in Table 1 were used in the calculations.

**4.2.2. Hamon and Vidal (1986) dataset**

Hamon and Vidal [56] reported experimental imbibition data of synthetic porous samples made of aluminum silicate. Filtered degassed water and purified refined oil were used as the wetting and non-wetting fluids, respectively. The samples analyzed in this study were subjected to the following boundary conditions during the experiments: both ends open, lower face open, and core fully immersed in water [56]. We used the reported fluid and rock core parameters from the counter-current imbibition experiments in our dimensionless calculations.

**4.2.3. Zhou et al. (2002) dataset**

Zhou et al. [55] conducted imbibition experiments on two diatomite samples from Grefco quarry, CA (Table 1). The samples were prepared as cylindrical cores and subjected to imbibition to determine fluid displacement mechanisms during counter-current SI. The experiments were conducted for both co-current and counter-current SI. Those authors periodically collected images to record oil recovery as a function of time. Water was used as the wetting phase, and n-decane, blandol, and air were used as non-wetting phases. The reported oil/air recovery versus imbibition time data were used to calculate the dimensionless time.

**4.2.3. Fischer et al. (2006) dataset**

Fischer et al. [54] reported experimental imbibition data for ten cylindrical Berea sandstones. They rinsed and oven-dried the cores for two days before carrying out counter-current imbibition experiments under different boundary conditions, including one end open, radial two ends closed, and all faces open. Refined oil was used as the non-wetting phase, and glycerol mixed with brine was used as the wetting-phase in the SI experiments. Oil production as a function of time was measured for the samples.

**4.2.4. Bourblaux and Kalaydjian (1990) dataset**

Bourblaux and Kalaydjian [57] reported experimental imbibition data for two fine, well-sorted Triassic sandstone samples from Vosges region, France. They cut the samples into a parallel-piped shape with rectangular cross sections of 61 x 21 mm and a length of 290 mm. The authors used brine and refined oil as the wetting and non-wetting fluids for the SI experiments. The blocks were subjected to both co-current and counter-current imbibition tests, and the oil recovery was measured and reported as a function of time.

**4.2.5. Babadagli and Hatiboglu (2007) dataset**

Babadagli and Hatiboglu [58] also reported imbibition data for six Berea sandstones, which were prepared as cylindrical core plugs with three different diameters of 2, 4, and 6 inches. The boundary condition used was one side open to imbibition, and the experiments were performed by positioning the cores horizontally and vertically. They conducted air-water imbibition experiments, using water at temperatures of 90°C and 20°C to investigate the effect of temperature on imbibition. Gas recovery was reported as a function of time for the experiments.

The values of $A$ needed to calculate the dimensionless time for the above-mentioned datasets were reported by Schmid and Geiger [1]. Other salient properties of the SI measurements used to evaluate the proposed scaling analysis in this study, Eq. (15), are reported in Table 1.

Table 1. Physical properties of the data used in scaling, $\alpha$ values, and data sources.

| Reference | Sample | $L_c$ (cm) | $k$ (mD) | $\phi$ (-) | $\mu_w$ (Pa.s) | $\mu_{nw}$ (Pa.s) | $S_i$ (-) | $A$ (m/$\sqrt{s}$) | $\alpha$ (-) |
|---|---|---|---|---|---|---|---|---|---|
| Zhang et al. (1996) | BA31 | 13.87 | 907.1 | 0.214 | $9.67 \times 10^{-4}$ | 0.03782 | 0 | $4.65 \times 10^{-6}$ | 1.18 |
| | BC13 | 4.99 | 503.6 | 0.209 | $9.67 \times 10^{-4}$ | 0.03782 | 0 | $7.05 \times 10^{-6}$ | 1.18 |
| | BC21 | 6.09 | 481.9 | 0.213 | $9.67 \times 10^{-4}$ | 0.00398 | 0 | $1.25 \times 10^{-5}$ | 0.95 |
| | BC22 | 5.68 | 496.8 | 0.208 | $9.67 \times 10^{-4}$ | 0.15630 | 0 | $4.46 \times 10^{-6}$ | 1.20 |
| | BD14 | 1.35 | 518.9 | 0.218 | $9.67 \times 10^{-4}$ | 0.03782 | 0 | $7.34 \times 10^{-6}$ | 1.12 |
| | BD15 | 1.35 | 523.8 | 0.214 | $9.67 \times 10^{-4}$ | 0.00398 | 0 | $1.28 \times 10^{-5}$ | 1.14 |
| | BD18 | 1.35 | 509.7 | 0.218 | $9.67 \times 10^{-4}$ | 0.15630 | 0 | $4.65 \times 10^{-6}$ | 1.13 |
| Haman and Vidal (1986) | A10 | 9.7 | 4000 | 0.472 | 0.001 | 0.0115 | 0.189 | $3.36 \times 10^{-5}$ | 1.07 |
| | A10-20 | 19.7 | 3430 | 0.453 | 0.001 | 0.0115 | 0.187 | $3.10 \times 10^{-5}$ | 1.07 |
| | A10-30 | 30.0 | 3830 | 0.453 | 0.001 | 0.0115 | 0.151 | $3.18 \times 10^{-5}$ | 1.09 |
| | A10-40 | 40.0 | 3550 | 0.478 | 0.001 | 0.0115 | 0.172 | $3.33 \times 10^{-5}$ | 1.07 |
| | A10-85 | 84.7 | 3000 | 0.478 | 0.001 | 0.0115 | 0.164 | $3.13 \times 10^{-5}$ | 0.90 |
| | A10VI-20 | 19.8 | 3200 | 0.456 | 0.001 | 0.0115 | 0.164 | $3.17 \times 10^{-5}$ | 1.21 |
| | A10X-20 | 20.0 | 2300 | 0.458 | 0.001 | 0.0115 | 0.132 | $2.92 \times 10^{-5}$ | 1.54 |
| Zhou et al. (2002) | Core-4 | 9.5 | 2.5 | 0.78 | 0.001 | $8.4 \times 10^{-4}$ | 0 | $2.72 \times 10^{-5}$ | 1.00 |
| | Core-5 | 9.5 | 6.0 | 0.68 | 0.001 | $8.4 \times 10^{-4}$ | 0 | $2.86 \times 10^{-5}$ | 1.03 |
| Fischer et al. (2006) | EV6-22 | 7.18 | 109.2 | 0.18 | 0.495 | 0.0039 | 0 | $4.29 \times 10^{-7}$ | 1.01 |
| | EV6-18 | 7.62 | 140.0 | 0.181 | 0.001 | 0.063 | 0 | $3.74 \times 10^{-6}$ | 1.04 |
| | EV6-21 | 7.7 | 107.3 | 0.187 | 0.0278 | 0.0039 | 0 | $1.86 \times 10^{-6}$ | 0.99 |
| | EV6-13 | 7.75 | 113.2 | 0.187 | 0.001 | 0.0039 | 0 | $7.48 \times 10^{-6}$ | 1.00 |
| | EV6-14 | 7.66 | 127.2 | 0.178 | 0.0041 | 0.0039 | 0 | $4.45 \times 10^{-6}$ | 1.00 |
| | EV6-20 | 7.52 | 132.9 | 0.181 | 0.0041 | 0.0633 | 0 | $2.45 \times 10^{-6}$ | 1.03 |
| | EV6-16 | 7.78 | 136.8 | 0.181 | 0.0278 | 0.0633 | 0 | $1.36 \times 10^{-6}$ | 1.00 |
| | EV6-23 | 7.36 | 132.1 | 0.179 | 0.0977 | 0.0039 | 0 | $9.92 \times 10^{-7}$ | 1.00 |
| | EV6-15 | 7.3 | 107.0 | 0.183 | 0.4946 | 0.0633 | 0 | $3.93 \times 10^{-7}$ | 0.99 |
| | EV6-17 | 7.54 | 128.1 | 0.19 | 0.0977 | 0.0633 | 0 | $8.59 \times 10^{-7}$ | 1.00 |
| Bourboaux and Kalaydjian (1990) | GVB-3 | 29.0 | 124.0 | 0.233 | 0.0012 | 0.0015 | 0.4 | $1.24 \times 10^{-5}$ | 1.02 |
| | GVB-4 | 14.5 | 118.0 | 0.233 | 0.0012 | 0.0015 | 0.411 | $9.67 \times 10^{-6}$ | 1.03 |
| Babadagli and Hatiboglu (2007) | F-11 | 10.16 | 500.0 | 0.21 | 0.001 | $1.8 \times 10^{-5}$ | 0 | $2.75 \times 10^{-5}$ | 1.01 |
| | F-12 | 15.24 | 500.0 | 0.21 | 0.001 | $1.8 \times 10^{-5}$ | 0 | $2.81 \times 10^{-5}$ | 1.06 |
| | F-14 | 10.16 | 500.0 | 0.21 | 0.001 | $1.8 \times 10^{-5}$ | 0 | $1.95 \times 10^{-5}$ | 1.04 |
| | F-16 | 5.08 | 500.0 | 0.21 | 0.001 | $1.8 \times 10^{-5}$ | 0 | $2.29 \times 10^{-5}$ | 0.88 |
| | F-17 | 10.16 | 500.0 | 0.21 | 0.001 | $1.8 \times 10^{-5}$ | 0 | $1.60 \times 10^{-5}$ | 1.01 |

| | F-18 | 15.24 | 500.0 | 0.21 | 0.001 | $1.8 \times 10^{-5}$ | 0 | $2.07 \times 10^{-5}$ | 0.99 |

### 4.3. Spontaneous imbibition in individual rough-wall fractures

We also investigated SI in rough-wall fractures. The experimental data used in this study are from Perfect et al. [59], who measured SI in three fractured Crossville sandstones and three Mancos shales. The porosity, fracture aperture, and fracture surface fractal dimensions were determined during the experiments. Perfect et al. [59] reported average porosity values of 5.85% and 5.59%, and average aperture sizes of 92 and 104 $\mu m$, respectively, for the Crossville sandstones and Mancos shales. In both rock types, they found the average surface fractal dimension to be 2.16, which falls in the typical range of 2 to 3 reported in the literature [60]. The SI experiments were carried out on the oven-dried cores and imaged during the SI experiments using a dynamic neutron radiography technique. Perfect et al. [59] reported the heights of the wetting front versus time (see their Figure 4). In this study, we fit the non-Boltzmann equation, $x = \lambda t^{\frac{\alpha}{2}}$, to the height-time data to optimize the exponent $\alpha$.

## 5. Results and Discussion

In this section, we first present the results of modeling the counter-current SI using either the GFFT, Eq. (4), or the time-fractal Richards' equation, Eq. (5), and discuss them in detail. We then demonstrate how the proposed generalized theoretical framework with variable $\alpha$ results in a much better collapse of the SI data.

### 5.1. Modeling counter-current SI
#### 5.1.1. Ferguson and Gardner (1963) dataset

As stated earlier, in this dataset, diffusivity was an increasing function of water content (see Fig. 5 from Ferguson and Gardner [45]), consistent with the definition given in Eq. (6). We, therefore, fitted the time-fractal Richards' equation, Eq. (5), to the water saturation-time data reported by Ferguson and Gardner [45]. The results showed that the solution of the time-fractal Richards' equation fit the data well for $x$ = 3.5, 7.5, 11.5, 15.5, 19.5, 21.5, 23.5, and 27.5 cm. Fits for $x$ = 3.5, 7.5, 15.5, and 27.5 cm are shown in Fig. 1 as examples. For other $x$ values, we found similar results. We found $\alpha$ = 0.93, 0.90, 0.91, 0.93, 0.87, 0.89, 0.88, and 0.93 (0.87 $\leq \alpha \leq$ 0.93) at $x$ = 3.5, 7.5, 11.5, 15.5, 19.5, 21.5, 23.5, and 27.5 cm, respectively. The results indicated that fluid flow in the studied soils did not follow the traditional Boltzmann transformation ($x = \lambda t^{1/2}$) in which $\alpha$ is fixed at 1.

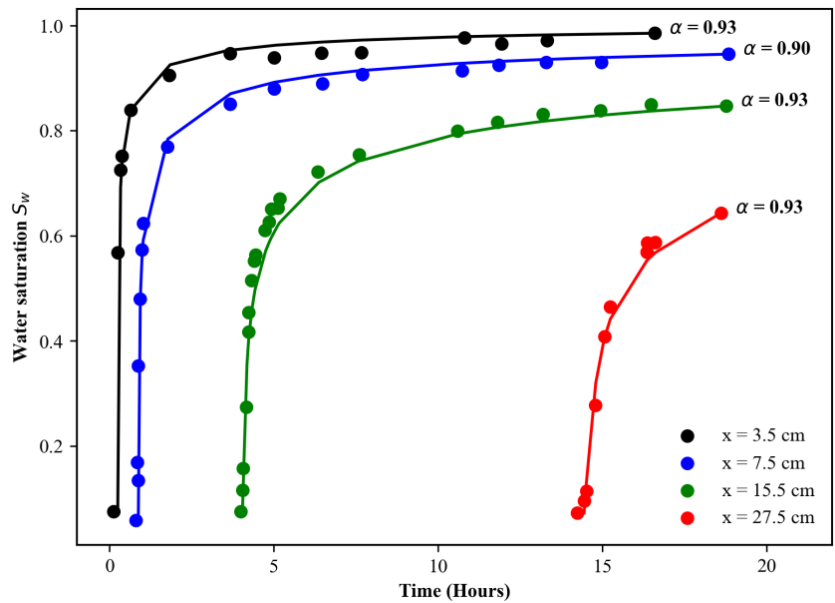

Figure 1. Water saturation, $S_w$, versus time, $t$, for four experiments at $x$ = 3.5, 5.5, 15.5, and 27.5 cm as reported by Ferguson and Gardner [45]. Filled circles represent the experimental measurements, while the solid lines denote the fitted time-fractal Richards'

equation, Eq. (5). The optimized $\alpha$ exponent in the non-Boltzmann transformation is shown next to each experiment.

As shown in Fig. 1, we found $\alpha$ = 0.93, 0.90, 0.93 and 0.93, with an average value of 0.921, for $x$ = 3.5, 7.5, 15.5 and 27.5 cm, respectively. This average value ($\alpha = 0.921$) is very close to 0.91 reported by Pachepsky et al. [42] who solved Eq. (5) using a different approach (see Eqs. (6)-(17) in Pachepsky et al. [42]). Pachepsky et al. [42] reported the average $\alpha$ = 0.91 by collapsing all four experiments, while we fit Eq. (5) individually to each experiment, optimize its value independently and then averaged all four values.

### 5.1.2. El Abd and Milczarek (2004) dataset

In this dataset, following Sun et al. [43], we used the power-law water diffusivity model, fit the solution of time-fractal Richards' equation (Eq. (5)) to the data, and optimized the exponent $\alpha$. Fig. 2 presents the fit of the solution of Eq. (5) to the fired-brick clay and siliceous brick experiments from El Abd and Milczarek [46]. As can be observed, the time-fractal Richards' equation characterized the wetting fronts reasonably well in most cases.

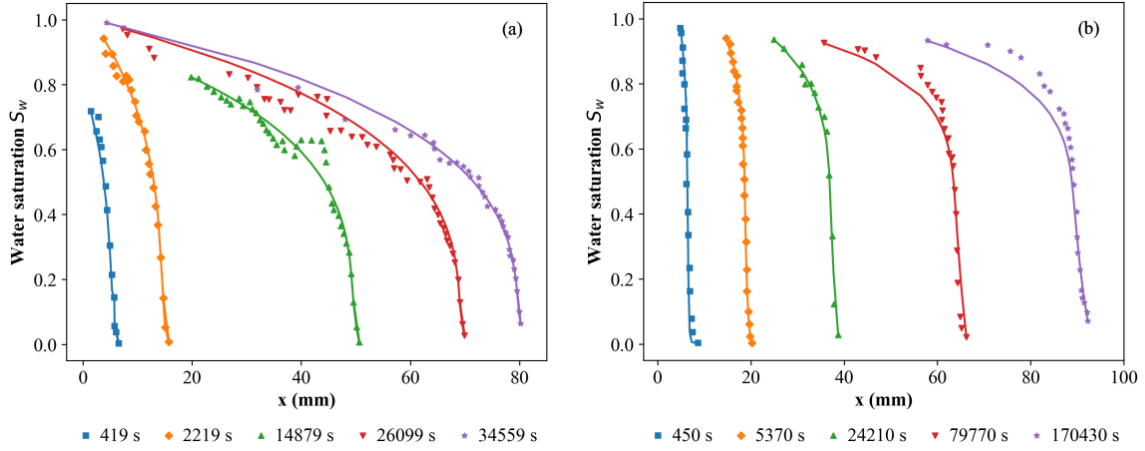

Figure 2. Water saturation, $S_w$, versus distance, $x$, at different times for the (a) fired-clay brick and (b) siliceous brick samples from El Abd and Milczarek [46]. Filled circles represent the experimental measurements, while the solid lines denote the fitted time-fractal Richards' equation (Eq. (5)).

By fitting the solution of Eq. (5) to each experiment, we optimized the non-Boltzmann scaling exponent and found $\alpha$ = 1.07, 1.07, 1.21, 1.14, and 1.14 at $t$ = 419, 2219, 14879, 26099, and 34559 seconds, respectively, for the fired-clay brick sample (Fig. 2a). The average value of 1.13 is near 1.24 reported by Sun et al. [43], who used a single exponent to characterize all experiments. Another reason for the discrepancy in the $\alpha$ values (1.13 vs. 1.24) could be due to differences in the approach used to solve the time-fractal Richards' equation in this study compared to Sun et al. [43]. For the siliceous brick sample, we found smaller $\alpha$ values compared to the fired-clay brick sample. Specifically, our fitting analysis resulted in $\alpha$ = 0.85. 0.85, 0.90, 0.93, and 0.90 at $t$ = 450, 5370, 24210, 79770, and 170430 seconds (Fig. 2b), respectively, with an average value of 0.89, which is very close to the 0.86 reported by Sun et al. [43].

### 5.1.3. Qin et al. (2022) dataset

In Fig. 3, we present the fit of the solution of GFFT, Eq. (4), to the numerical stimulations of SI for the Estaillades carbonate reported by Qin et al. [47]. We found $\alpha$ = 0.90, 0.83, 0.90, and 1.08 for contact angles of 20°, 40°, 60°, and 80°, respectively. The results clearly show the dependency of $\alpha$ on wettability, similar to the findings of Bakhshian et al. [61], who conducted color gradient lattice-Boltzmann simulations on water- and intermediate-wet rock images of Tuscaloosa sandstone. They reported that wettability was a major factor causing scaling exponents different from 0.5 in the Boltzmann transformation (where $\alpha$ = 1).

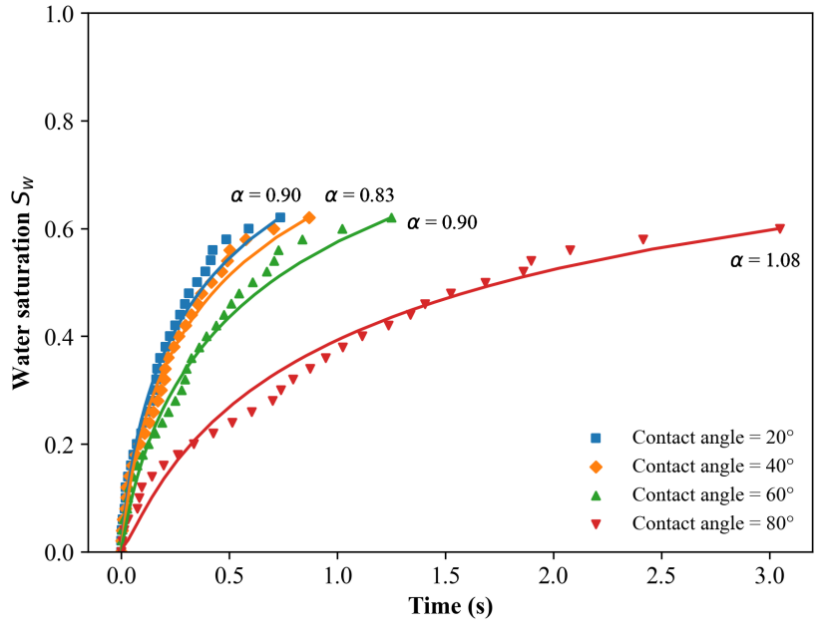

Figure 3. Water saturation, $S_w$, versus time, $t$, for four sets of simulations corresponding to contact angles of 20°, 40°, 60°, and 80° reported by Qin et al. [47] for the Estaillades carbonate. Filled markers represent the numerical simulations, while the solid lines denote

the fitted GFFT, Eq. (4). The optimized $\alpha$ exponent in the non-Boltzmann transformation is shown next to each simulation.

Generally speaking, we found an increasing trend between the non-Boltzmann scaling exponent and the contact angle. However, further investigations are still required to study the effect of wettability on the exponent $\alpha$ more comprehensively, using a wider range of contact angles as well as in heterogeneously wet media. The $\alpha$ values less than or greater than 1 found in this study clearly demonstrate the application of the GFFT, Eq. (4). Specifically, our results demonstrated that the non-Boltzmann scaling exponent $\alpha$ depends not only on the heterogeneity of pore space but also on fluid-solid properties, such as the contact angle.

### 5.1.4. Meng et al. (2022) dataset

Results of fitting the GFFT, Eq. (4), to three sets of experimental SI data reported in quartz sand packs by Meng et al. [51] are presented in Fig. 4. We found $\alpha$ = 1.05, 0.85, and 1.10 for dynamic non-wetting viscosities $\mu_{nw}$ = 2.8 mPa s (kerosene), 25.6 mPa s (white oil-15), and 103.4 mPa s (white oil-32), respectively. Although the trend between $\alpha$ and $\mu_{nw}$ is non-monotonic, our results clearly show the dependency of $\alpha$ on the dynamic viscosity of the non-wetting phase. Diao et al. [62] also reported that mixed wettability or different viscosity ratios affected SI in heterogenous porous media, using numerical stimulations with the quasi-three-dimensional color-gradient lattice-Boltzmann method of Liu et al. [17]. Based on the relationship between wetting saturation and imbibition time, Diao et al. [62] showed that the scaling exponent deviated from the Boltzmann transformation, where $\alpha = 1$. This is further supported by our results, using the GFFT to characterize SI, which shows scaling exponents different from 0.5 in the Boltzmann transformation (where $\alpha = 1$).

To analyze the experiments of Meng et al. [51] and determine the non-Boltzmann scaling exponent $\alpha$, we had to use the estimated capillary pressure and relative permeability curves, as such data were not reported by Meng et al. [51]. More accurate $\alpha$ values and SI characterization by the GFFT should be expected when such input data are available.

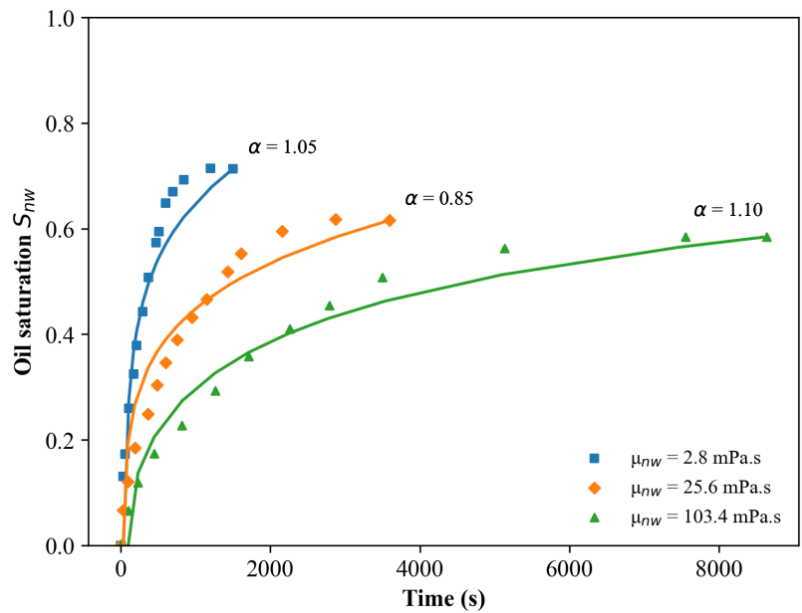

Figure 4. Oil saturation, $S_{nw}$, versus time, $t$, for three experiments corresponding to non-wetting-phase viscosities $\mu_{nw}$ = 2.8 mPa s (kerosene), 25.6 mPa s (white oil-15), and 103.4 mPa s (white oil-32), as reported by Meng et al. [51] for the quartz sand packs. Filled markers represent the experimental measurements, while the solid lines denote the fitted GFFT, Eq. (4). The optimized $\alpha$ exponent in the non-Boltzmann transformation is shown next to each experiment.

## 5.2. Scaling SI data across various porous media

Results of the SI scaling analyses are presented in Fig. 5. The unscaled SI data shown in Fig. 5a indicate a wide range of data diversity and scatter in the measurements for various types of porous media, experimental conditions, and water-air and water-oil mixtures. In Fig. 5b, we show the scaled SI data using the Schmid and Geiger [1] approach, a special case of our GFFT model with $\alpha = 1$. As can be observed, although some experiments fall onto a curve, there is still scatter in the data. It is important to note that Fig. 5b in this work differs slightly from Fig. 4b in Schmid and Geiger [1], possibly because we collected SI recovery data more extensively from the same literature as Schmid and Geiger [1]. Fig. 5c exhibits the results of our proposed scaling approach based on the GFFT in combination with the non-Boltzmann transformation. The scaled SI data nearly collapsed onto a single universal curve, with reasonable $\alpha$ values ranged from 0.88 to 1.54 (Table 1). This collapse in the data confirms the practical application of the proposed scaling analysis. Note that to scale the SI data reported here, we used the $A$ values and other parameters like porosity ($\phi$) and characteristic length ($L_c$) reported by Schmid and Geiger [1]. They used estimated capillary pressure and relative permeability data to calculate diffusivity based on the analytical approach of McWhorter and Sunada [38]. We posit that if experimental capillary pressure and relative permeability data are used as direct inputs, the proposed scaling analysis can scale SI more accurately, as it minimizes errors.

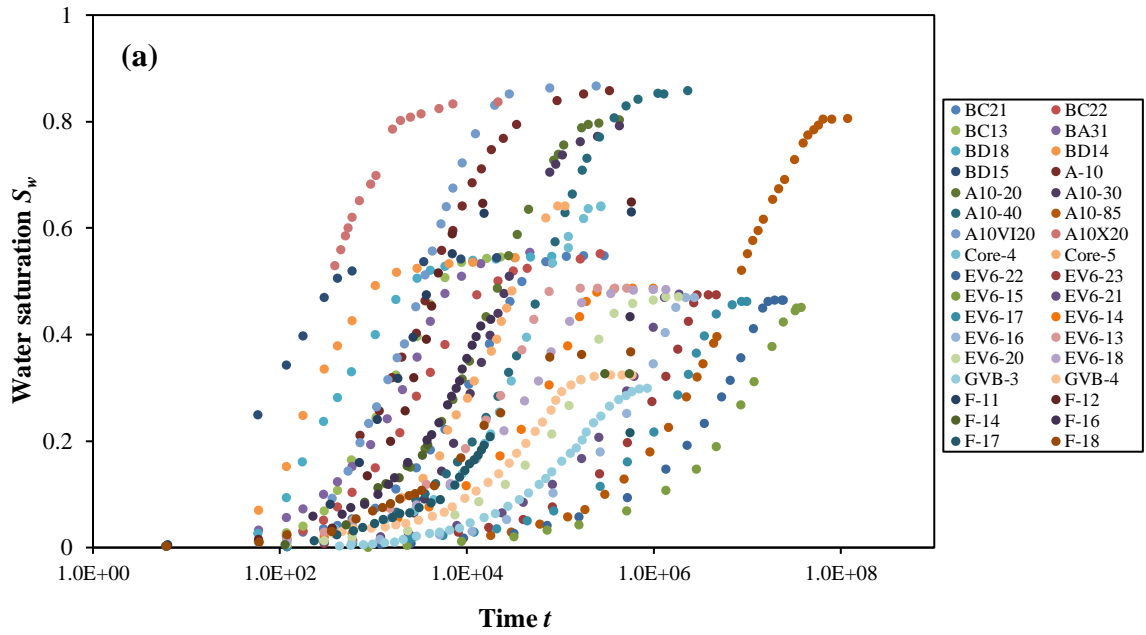

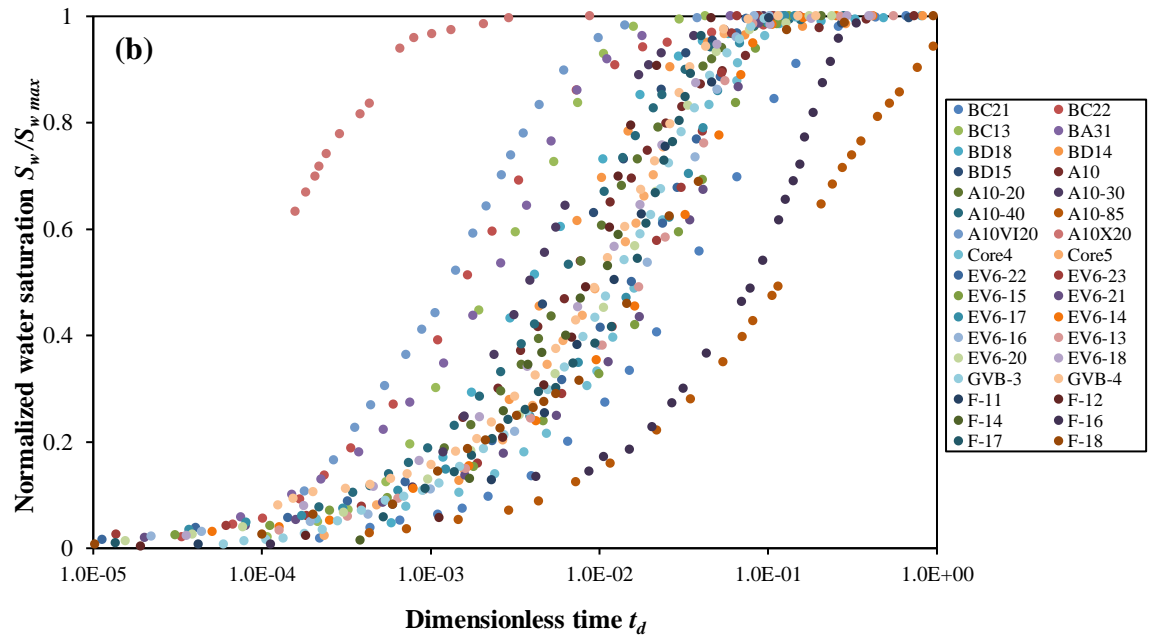

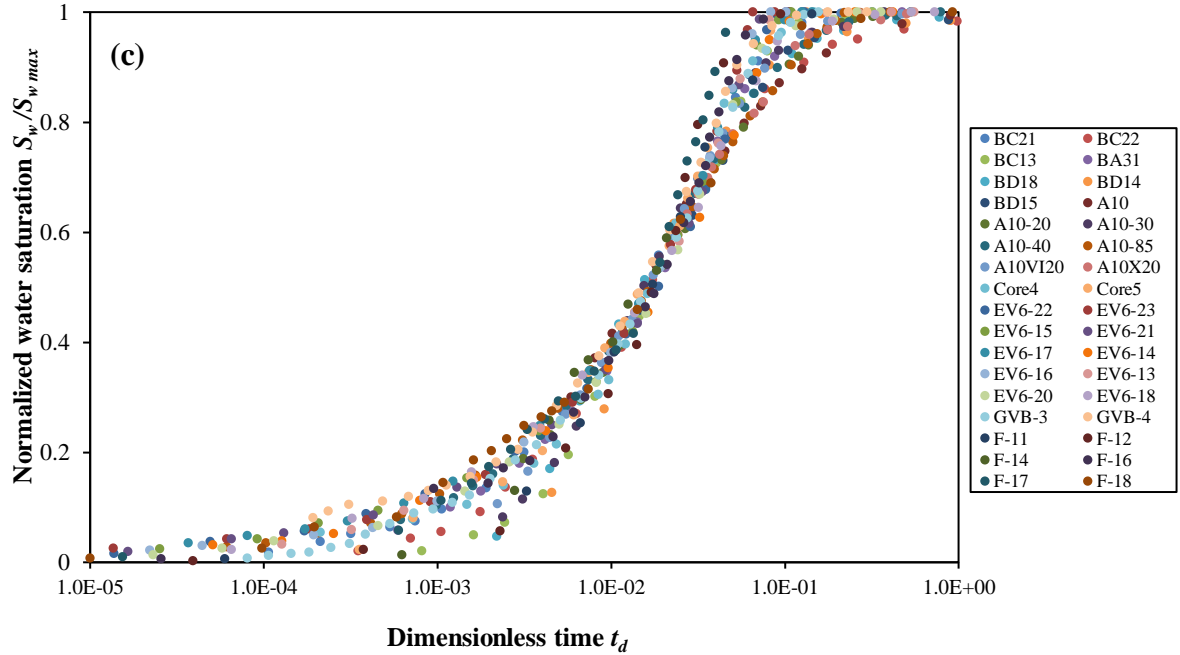

Figure 5. (a) Water saturation versus time collected from the literature; (b) Normalized water saturation against dimensionless time based on scaling analysis of Schmid and Geiger [1]; (c) Normalized water saturation against dimensionless time, Eq. (15), proposed in this study.

We also applied the proposed scaling approach to four SI sets of simulations on the Estaillades carbonate sample reported by Qin et al. [47], and the experimental and stimulation SI data on quartz sand pack samples by Meng et al. [51]. We again compared the scaling with the model introduced by Schmid and Geiger [1]. Fig. 6 compares the scaling of the imbibition curves with the Boltzmann approach and the non-Boltzmann approach for the SI data reported by Qin et al. [47] and Meng et al. [51]. The results showed that the data did not collapse well with $\alpha$ being fixed at 1, or in other words, using the previous approach, but it achieved a satisfactory collapse when we used our approach with varying $\alpha$ values. Additionally, in some cases, such as the data for contact angles 40º and 60º in Qin et al. [47], and the data for dynamic non-wetting viscosity of 2.8 mPa s in Meng et al. [51], we were able to find closer $\alpha$ values to those obtained by fitting the GFFT to the imbibition curves. This result provides solid evidence that using the GFFT in combination with the non-Boltzmann transformation in modeling two phase flow and scaling SI predicts flow behavior and captures heterogeneity more accurately than the classical Richards' equation and Boltzmann transformation, which restricts the flexibility of the imbibition exponent. Moreover, this observation confirms that the $\alpha$ value depends on both the contact angle and the dynamic viscosity of the fluids used.

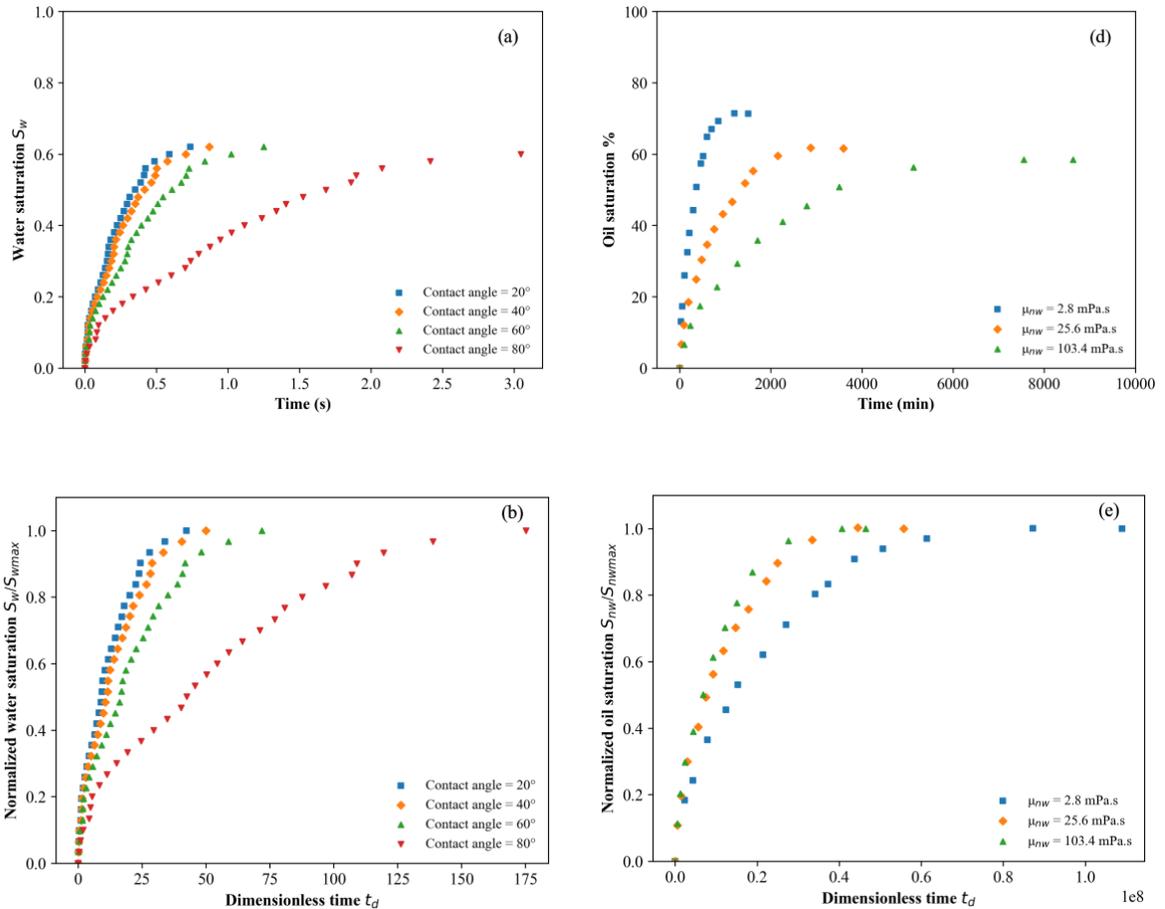

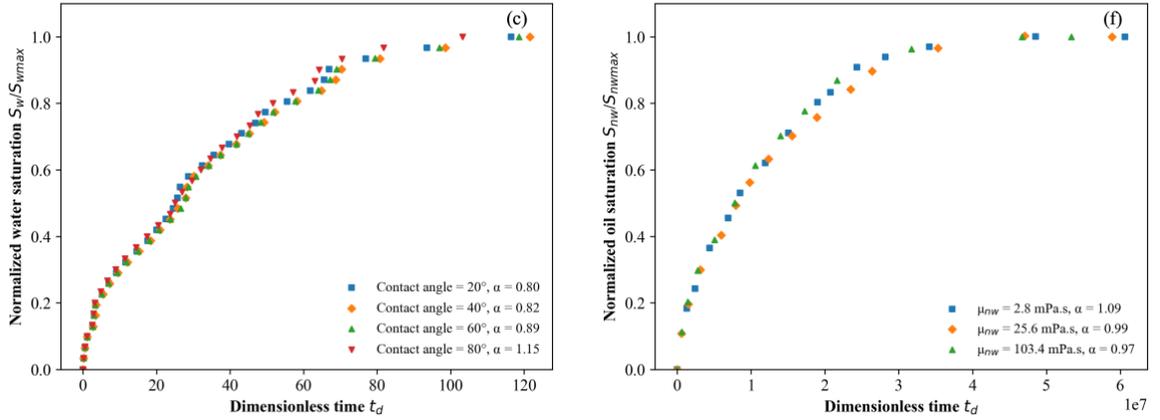

Figure 6. Scaling of SI data reported by Qin et al. [47] and Meng et al. [51] using two approaches: (a) Water saturation versus time collected from Qin et al. [47]; (b) Normalized water saturation against dimensionless time based on the scaling analysis of Schmid and Geiger [1] for data reported in (a); (c) Normalized water saturation against dimensionless time based on Eq. (15), proposed in this study, for data reported in (a); (d) Water saturation versus time collected from Meng et al. [51]; (e) Normalized water saturation against dimensionless time based on the scaling analysis of Schmid and Geiger [1] for data reported in (d); (f) Normalized water saturation against dimensionless time based on Eq. (15), proposed in this study, for data reported in (d).

Based on Figs. 6c and 6f, it is evident that the non-Boltzmann approach used in this study collapses SI data into a universal curve with less scattering than the previous model used by Schmid and Geiger [1]. The $\alpha$ values which yield a satisfactory collapse of the curves are reported in the legend of Figs. 6c and 6f.

## 5.3. Non-Boltzmann scaling in individual rough-wall fractures

We also investigated the variation of $\alpha$ in rough-wall fractures during imbibition by fitting the non-Boltzmann transformation equation, $x = \lambda t^{\frac{\alpha}{2}}$, to the wetting front height against the imbibition time data reported by Perfect et al. [59]. The fits shown in Fig. 7 are reasonable, with $R^2 > 0.95$. We found $\alpha = 1.15$, 1.24, and 1.07, with an average value of 1.15, for the three Crossville sandstones. For the three Mancos shale samples, we found $\alpha = 0.87$, 0.92, and 0.98, with an average value of 0.92. Results showed exponents different from 0.5, meaning that Boltzmann scaling does not necessarily scale time and height in rough-wall fractures. Although both sandstone and shale samples studied here had the same surface fractal dimension (equal to 2.16), the average $\alpha$ value for the Crossville sandstone was different from that for the Mancos shale. This indicates the minor impact of surface roughness on the non-Boltzmann scaling exponent $\alpha$. Interestingly, we found an average $\alpha = 1.15$ for the Crossville sandstone samples with an average aperture of 92 $\mu m$, while the average $\alpha = 0.92$ for the Mancos shale samples with an average aperture of 104 $\mu m$, implying an inverse trend between the exponent $\alpha$ and the average aperture size. Recently, Ning et al. [31] investigated the effect of the presence of micro-fractures within the shale rock matrix and found exponents different from 0.5. They argued that micro-fracture orientation (e.g., parallel or transverse) caused deviations from the Boltzmann scaling. However, further investigations using a broader range of fractures with various roughness and aperture sizes are still needed to study how the exponent $\alpha$ may vary within individual rough-wall fractures.

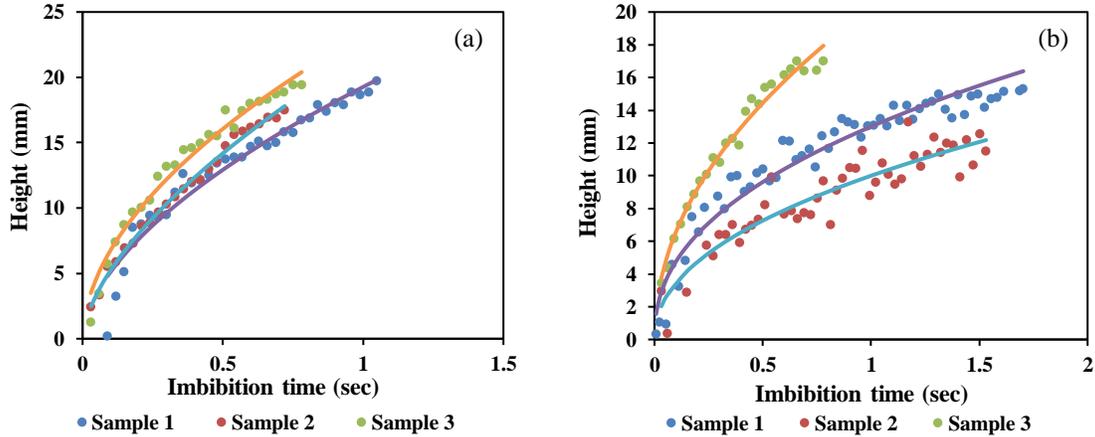

Figure 7. Experiments reported by Perfect et al. [59], shown by filled circles, and the fitted non-Boltzmann equation, shown by solid lines for (a) Crossville sandstones and (b) Mancos shale.

## 6. Conclusions

Modeling and scaling counter-current SI have been a long-standing challenge over the past several decades and have several practical applications, such as $CO_2$ sequestration. Although various SI models have been developed in the literature, they are often limited to a certain rock type or specific parameter scenario. In this study, we generalized fractional flow theory using concepts of time-fractal derivatives and non-Boltzmann transformation to model counter-current SI. We validated the new scaling model by analyzing more than 30 samples, including sandstones, diatomite, synthetic porous media, and carbonates with varying porosities, initial water contents, and dynamic viscosities, demonstrating the model's broad applicability across various porous media. The scaled data falls onto a universal curve, strongly indicating the model's suitability for scaling SI data. We suggest using measured experimental multiphase flow properties, namely capillary pressure and relative permeability, to further improve scaling. In fact, capturing these properties is vital for calculating $t_d$ and accurately scaling the data using the model.

This work also reveals that the use of the traditional Richards' equation, which implies that the wetting front in a medium follows Boltzmann scaling with the horizontal travel distance proportional to the square root of time, is not applicable or accurate for most of the porous media and soil data examined. Instead, we emphasize and demonstrate that using a GFFT approach better captures the fractal nature of heterogenous porous media and unsaturated soil. The solution of GFFT shows non-Boltzmann scaling with the time exponent varying around 0.5. Fitting the GFFT solution and the $\alpha$ values to the wide variety of collated datasets reveals that the Boltzmann scaling time exponent significantly varies from 0.5.

The analysis suggests that the variation in the $\alpha$ value can be attributed to differences in the contact angle. The water content curves with four contact angles collapse onto a single curve with varying $\alpha$ values. We also show that the obtained $\alpha$ values closely match the GFFT fitting. However, the effect of contact angle on $\alpha$ should be further studied. Additionally, the $\alpha$ value depends on viscosity ratios, as demonstrated by our prediction of fluid fronts using the GFFT approach, yielding variable $\alpha$ values. It is evident that the GFFT approach supports previous work on imbibition time exponents, which characterized deviations of the imbibition exponent from 0.5 (in this work, we refer to the $\alpha$ value in the imbibition exponent, with $\alpha/2$ as the imbibition exponent), based on factors like pore connectivity, pore structure, fracture networks inside porous media, and wettability. This study lays the foundation for further investigate into the relationship of $\alpha$ with contact angle, viscosity ratio, lithologies, and pore structure across samples. It is also noteworthy that the GFFT approach is highly sensitive to the diffusivity function used as an input. Diffusivity depends on primary inputs like capillary pressure and relative permeability, and the use of experimental data for these inputs is highly recommended to maintain accuracy and minimize errors in the outputs. Using this new approach, we expect to more accurately predict the rate and volume of fluid that can spontaneously imbibe into a porous medium, provided characteristics such as porosity, permeability, initial and maximum saturations, viscosity, and wettability are known.

**Acknowledgment**

The authors express their gratitude to Emeritus Prof. David B. McWhorter for fruitful discussions on fractional flow theory. BG acknowledges University of Texas at Arlington for supports through faculty startup fund and the STARs award.

**Notation**

| | |
|---|---|
| $A$ | Scaling constant |
| $\alpha$ | Non-Boltzmann variable in the imbibition exponent |
| $S_w$ | Water saturation |
| $t$ | Time |
| $f$ | Fractional flow function for viscous flow |
| $x$ | Distance |
| $D$ | Two-phase flow diffusivity |
| $P_c$ | Capillary pressure |
| $\phi$ | Porosity |

| Symbol | Description |
|---|---|
| $q_t$ | Total flux |
| $q_w$ | Wetting phase flux |
| $q_{nw}$ | Non-wetting phase flux |
| $\lambda$ | Lambda |
| $D_\alpha$ | Fractal diffusivity |
| $t^\alpha$ | Fractal time |
| $\theta$ | Water content |
| $S_i$ | Initial water saturation |
| $S_o$ | Final water saturation |
| $S_{wmax}$ | Maximum water saturation |
| $V_b$ | Bulk volume of the sample |
| $L_c$ | Characteristic length |
| $a_i$ | Area of the open face |
| $l_{ai}$ | Distance traveled by the imbibition front toward the no-flow boundary |
| $t_d$ | Dimensionless time |
| $Q_w$ | Cumulative imbibition |
| $k$ | Permeability |
| $\mu_w$ | Dynamic viscosity of the wetting phase |
| $\mu_{nw}$ | Dynamic viscosity of the non-wetting phase |
| $k_w$ | Relative permeability of the wetting phase |
| $k_{nw}$ | Relative permeability of the non-wetting phase |

## Appendix A. Derivation of the dimensionless time $t_d$

This appendix explains the origin of the exponent $2/\alpha$ in the definition of dimensionless time $t_d$ in Eq. (13). The non-Boltzmann scaling describes the wetting-phase front as:

$$x(t) = \lambda(S_w) t^{\alpha/2}, \tag{A1}$$

where $\lambda$ is a function of the wetting-phase saturation. Eq. (A1) generalizes the Boltzmann scaling for the standard FFT model, Eq. (3):

$$x(t) = \lambda(S_w) t^{1/2}. \tag{A2}$$

The form of $\lambda(S_w)$ was derived by Schmid [39] (see her equation (2.29)):

$$\lambda(S_w) = \frac{2A(1-fR)}{\phi} \frac{dF(S_w)}{dS_w}, \tag{A3}$$

where the scaling constant $A$ is defined by Eq. (14a), $f$ is the fractional flow function, $R = q_t/q_0$, and $F = \frac{q_w/q_0 - fR}{1-fR}$ is calculated using Eq. (14b). For the case of counter-current SI considered in this study, equation (A3) can be simplified as follows:

$$\lambda(S_w) = \frac{2A}{\phi}. \tag{A4}$$

Combining Eq. (A1) with Eq. (A4) yields an effective wetting-front flux:

$$q_w = \frac{dx(t)}{dt} = \frac{\alpha A}{\phi} t^{\frac{\alpha}{2}-1}, \tag{A5}$$

and the cumulative wetting-phase flux by time $t$ is approximated as:

$$Q_w(t) = \int_0^t q_w(s)\, ds = \frac{2A}{\phi} t^{\alpha/2}. \tag{A6}$$

This shows that the non-Boltzmann scaling rate of the wetting-phase front also defines the cumulative wetting-phase flux in counter-current SI. The dimensionless time $t_d$ corresponding to clock time $t$ can then be defined as the scaled ratio of cumulative flux to the medium's characteristic length $L_c$ (defined by Eq. (15)), leading to:

$$t_d(t) = \left[\frac{Q_w(t)}{\sqrt{\varepsilon}L_c}\right]^{\frac{2}{\alpha}} = \left(\frac{2A}{\phi\sqrt{\varepsilon}L_c}\right)^{\frac{2}{\alpha}} t. \tag{A7}$$

The term $\sqrt{\varepsilon}$ is added here for unit consistency. Notably, the fractal index $\alpha$, which is related to the medium's fractal dimension, may influence the wetting-phase travel path, similar to the effect of tortuosity. A smaller $\alpha$ in the GFFT equation captures slower wetting-front movement, analogous to increased tortuosity or a longer travel path. This hypothesis may explain why the exponent in the dimensionless travel time is the inverse of $\alpha$. Further research is needed to test this hypothesis and explore alternative formulations of Eq. (A7).